\documentclass[aps,prb,superscriptaddress,longbibliography, twocolumn]{revtex4-1}

\usepackage{amsmath}
\usepackage{todonotes}
\usepackage[colorlinks=true, allcolors=black, citecolor=blue, linkcolor=blue, urlcolor=blue]{hyperref}
\usepackage{microtype}
\usepackage[capitalize]{cleveref}

\newcommand{\gc}{\check{g}}

\newcommand{\gA}{\hat{g}^{A}}

\newcommand{\gR}{\hat{g}^{R}}

\newcommand{\gK}{\hat{g}^{K}}

\newcommand{\ds}{\partial_{s}}

\newcommand{\dst}{\tilde{\partial}_{s}}

\newcommand{\Dtilde}{\tilde{D}}
\newcommand{\dtilde}{\tilde{\partial}}

\newcommand{\comm}[2]{\left[ #1, #2 \right]}

\newcommand{\hhat}{\hat{h}}
\newcommand{\that}{\hat{\tau}}

\newcommand{\That}{\hat{{T}}(s)}
\newcommand{\Nhat}{\hat{{N}}(s)}
\newcommand{\Bhat}{\hat{{B}}(s)}

\newcommand{\mtupper}{{G}^{\mu\nu}}

\DeclareMathOperator{\Tr}{Tr}

\newcommand{\Nt}{\Tilde{N}}
\newcommand{\gt}{\Tilde{\gamma}}

\begin{document}
\title{Non-constant geometric curvature for tailored spin-orbit coupling and chirality in superconductor-magnet heterostructures}

\author{Alv Johan Skarpeid}
\affiliation{Center for Quantum Spintronics, Department of Physics, NTNU, Norwegian University of Science and Technology, NO-7491 Trondheim, Norway}
\affiliation{Department of Physics, Centre for Materials Science and Nanotechnology, University of Oslo, NO-0316 Oslo, Norway}
\author{Henning G. Hugdal}
\affiliation{Center for Quantum Spintronics, Department of Physics, NTNU, Norwegian University of Science and Technology, NO-7491 Trondheim, Norway}
\author{Tancredi Salamone}
\affiliation{Center for Quantum Spintronics, Department of Physics, NTNU, Norwegian University of Science and Technology, NO-7491 Trondheim, Norway}
\author{Morten Amundsen}
\affiliation{Center for Quantum Spintronics, Department of Physics, NTNU, Norwegian University of Science and Technology, NO-7491 Trondheim, Norway}
\affiliation{Nordita, KTH Royal Institute of Technology and Stockholm University, Hannes Alfvéns väg 12, SE-106 91 Stockholm, Sweden}
\author{Sol H. Jacobsen}
\email[Corresponding author: ]{sol.jacobsen@ntnu.no}
\affiliation{Center for Quantum Spintronics, Department of Physics, NTNU, Norwegian University of Science and Technology, NO-7491 Trondheim, Norway}

\date{\today}

\begin{abstract}
We show that tailoring the geometric curvature profile of magnets can be used for bespoke design of an effective non-relativistic spin-orbit coupling, which may be used to control proximity effects if the magnet is coupled to a superconductor. We consider proximity-coupled one-dimensional magnetic wires with variable curvatures, specifically three distinct shapes classified as J-, C-, and S-type. We demonstrate a chirality-dependent spin polarization of the superconducting correlations, and show the role of curvature in determining the ground state of mixed-chirality junctions. We speculate on how this may be implemented in novel device design, and include analysis of its usage in a spin-triplet SQUID.
\end{abstract}

\maketitle

\section{Introduction}
Combining the typically competing phases of superconductivity and magnetism provides fertile ground for uncovering fundamental physics, and is essential for advancing the field of superconducting spintronics \cite{Eschrig2011,Linder2015}. Resistance-free spin and charge transport in superconductors can give an energy advantage in spintronics, where high current densities are needed for novel information processing architectures. However, there are a limited number of combination mechanisms, and strong restrictions on the tailoring and control of these. Geometric curvature has recently emerged as providing a range of new freedoms for design and control \cite{Gentile2022,salamone2022curvature,Salamone2023}, and here we examine the implications of non-constant curvatures in magnets coupled to superconductors.

The most abundant superconductors are robust to impurities, with conventional, singlet, s-wave orbital correlations (Cooper pairs), where averaging over scattering events in momentum-space leaves a finite superconducting order parameter. However, singlet pairs are rapidly destroyed in a spin-polarized material such as a ferromagnet. To combine diffusive s-wave, singlet superconductivity with magnetism, we instead need to convert the singlet pairings into odd-frequency \cite{Linder2019} spin-polarized triplets. This is done via the proximity effect, where properties of adjacent materials merge through their interfacial barriers \cite{Bergeret2005,Buzdin2005,Lyuksyutov2005}. Traditionally, the conversion is manufactured via magnetic inhomogeneities, such as misaligned magnetic multilayers or an intrinsically helical magnetic lattice structure \cite{Bergeret2001,Robinson2010,Khaire2010}, or via intrinsic spin-orbit coupling (SOC) \cite{Bergeret2013,Bergeret2014,Jacobsen2015b}. However, manipulating individual magnetic multilayers can be challenging experimentally, and helical spin lattice textures are fixed, and cannot be varied or tailored. Moreover, intrinsic SOC is a relativistic effect, dependent on non-centrosymmetric crystal structures and/or spatially restricted interfacial symmetry breaking, so options for tailoring and controlling such systems are rather limited. In contrast, curvature in real-space (as opposed to the related field of band-structure curvature) can be a source of non-relativistic spin orbit effects in magnetic systems, which we have shown can be harnessed to tailor and control diffusive proximity effects for combining superconductivity and magnetism \cite{Salamone2021,salamone2022curvature,Salamone2023}. For instance, curvature can control the direction of charge current flow through a Josephson junction \cite{Salamone2021}, it can act as a probe of the quality of an uncompensated antiferromagnetic interface \cite{Salamone2023}, or even control the superconducting transition itself \cite{salamone2022curvature}. 

Research into the role of curvature has increased dramatically in recent years \cite{Gentile2022}, for example showing interesting effects in semiconductors \cite{Nagasawa2013,Gentile2015,Ying2016,Ying2017,Chang2017} and in superconductors \cite{Turner2010,Francica2020,Kutlin2020,Chou2021}. The curvature can for example promote topological edge states \cite{Gentile2015} and topological superconductivity \cite{Chou2021}. Curvature induces a quantum geometric potential \cite{Cantele2000,Aoki2001,Encinosa2003,Ortix2010}, and may induce a strain field producing a curvature-induced Rashba-type spin-orbit coupling \cite{Sheka2020, Jeong2011,Gentile2013,Wang2017}. If a material is magnetic, Rashba SOC leads to a chiral or extrinsic Dzyaloshinskii-Moriya interaction (DMI) and magnetic anisotropy \cite{Streubel2016,Sheka2021,Streubel2021,Imamura2004,Kim2013,Kundu2015,Gaididei2014}, and it is the curvature-controlled rotation of the spin orientation axis that is the source of non-relativistic spin-orbit effects in magnets \cite{Salamone2021,salamone2022curvature,Salamone2023}. Curvature-induced DMI can cause the appearance of chiral and topological spin textures of the effective magnetization in toroidal nanomagnets \cite{Vojkovic2017,Teixeira2019}, bent nanotubes \cite{Otalora2012,Almocidad2020}, curved surfaces \cite{Santos2013}, nanohelices \cite{Volkov2018,Pylypovskyi2020} and spherical shells \cite{Kravchuk2012,Gaididei2014,Sheka2015}. 

Materials design and fabrication of nanostructures with curved geometries has also been rapidly advancing to create ever smaller and more intricate nanoscale designs, from etching \cite{Schmidt2001,Cendula2009} and compressive buckling \cite{Xu2015}, to electron beam lithography \cite{Lewis2009,Burn2014,Volkov2019}, two-photon lithography \cite{Williams2018,Askey2020}, glancing angle deposition \cite{Dick2000} and focused electron beam induced deposition \cite{Sanz-Hernandez2020,Dobrovolskiy2021}. The curved designs can be manipulated in situ via strain, with dynamical control via photostriction, piezoelectrics, thermoelectric effects, tuning of the surface chemistry and more \cite{Kundys2015,Matzen2019,Guillemeney2022}.

\begin{figure*}[t!hbp]
    \includegraphics[width = \linewidth]{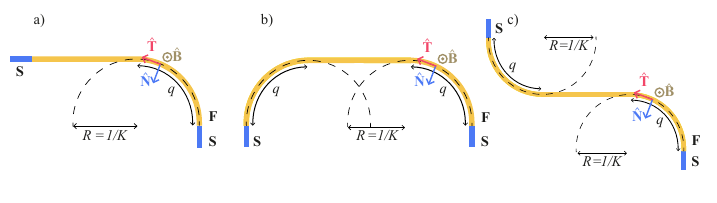}
    \caption{The (a) J-, (b) C- and (c) S-type superconductor-ferromagnet-superconductor (SFS) junction. The three orthonormal basis vectors \(\That, \Nhat\) and \(\Bhat\) are indicated in (a). Here, \(q\) is the parameter indicating the length of the curved section(s) and \(K > 0\) the curvature amplitude of the curved sections; a segment with curvature amplitude \(K\) follows a circular arc of radius \(R = 1/K\).}
    \label{fig:JSC_example} 
\end{figure*}

In this article, we investigate non-constant curvature in three classes of magnetic nanowires (J-, C- and S-type; see \cref{fig:JSC_example}), and demonstrate how such variable curvatures can influence the proximity effect and spintronic device design. We show a chirality-dependent spin polarization, investigate the ground state in mixed-chirality junctions, and we discuss this in the context of a spin-triplet SQUID design.

\section{Theory}\label{sec:Theory}

In Sec.~\ref{Sec:Usadel}, we introduce the quasiclassical Usadel equation of motion and relevant boundary conditions for diffusive spin transport, generalized to curved heterostructures. In Sec.~\ref{Sec:Non-const} we present the parameterisation of curves with non-constant curvature that define the classes of curves (J, C, and S) presented in \cref{fig:JSC_example}. The numerical approach for solving the spin transport equations via Riccati parameterisation is summarized in Sec.~\ref{Sec:Riccati}. 

\subsection{Quasiclassical formalism: Usadel equation in curvilinear coordinates}\label{Sec:Usadel}

We employ the quasiclassical formalism \cite{Belzig98, usadel1970} where observables are retrievable from a propagator $\gc$ in Keldysh \(\otimes\) Nambu (particle-hole) \(\otimes\) spin space, 
\begin{equation}
    \gc = 
    \begin{pmatrix}
        \gR & \gK \\
        0 & \gA 
    \end{pmatrix},
\end{equation}
where the superscripts $R$, $A$ and $K$ label the retarded, advanced and Keldysh components, respectively.
The elements are related by \(\gA = - \that_3 \hat{g}^{R\dagger} \that_3\) and \(\gK = \gR \hhat - \hhat \gA\), where we refer to \(\hhat\) as the \emph{distribution matrix} and \(\that_3 = \text{diag}(1,1,-1,-1)\).

The propagator $\gc$ obeys the Usadel equation \cite{usadel1970}, which in curvilinear coordinates is given as \cite{salamone2022curvature}
\begin{equation}
    \label{eq:usadel_curvilinear}
    D_{F} \mtupper \Dtilde_{\mu} \left(\gc \Dtilde_{\nu} \gc\right)
    + i \comm{\epsilon \that_{3} \otimes I_{2} + \check{\Sigma}}{\gc} = 0, 
\end{equation}
where $D_F$ is the diffusion coefficient, $\epsilon$ is the energy, \(\check{\Sigma}\) is the self-energy function, $G^{\mu\nu}$ is the metric tensor, and \(I_{2}\) is the \(2\times 2\) identity matrix in Keldysh space. 
The coordinate-gauge covariant derivatives are given as
\begin{equation}
    \label{eq:sgc_derivatives}
    \Dtilde_{\mu} v_{\nu} = \dtilde_{\mu} v_{\nu}  - \Gamma^{\gamma}_{\mu\nu} v_{\gamma}, 
\end{equation}
with Christoffel symbols
\begin{equation}
\Gamma_{\mu\nu}^{\gamma} =
\frac{1}{2} G^{\gamma\lambda}
\left[\partial_{\nu} G_{\mu \lambda} + \partial_{\mu}G_{\lambda\nu} - \partial_{\lambda}G_{\mu\nu}\right],
\end{equation}
and gauge-only covariant derivative $\dtilde_\mu v_\nu=\partial_\mu v_\nu-i[\hat{A}_\mu,v_\nu]$. The gauge field $\hat{A}_\mu=\rm{diag}(A_\mu,-A_\mu^*)$ will depend on e.g. intrinsic SOC contributions.

The Christoffel symbols depend on the choice of coordinate system, and we consider a planar space curve in real, 3-dimensional space, \(\boldsymbol{r}(s)\), at the center of a ferromagnetic nanowire. 
This space, and the nanowire, is parametrizable as \(\boldsymbol{R}(s, n, b) = \boldsymbol{r}(s) + n \Nhat + b \Bhat\), where we refer to \(s\), \(n\) and \(b\)
as the \emph{arclength, normal} and \emph{binormal coordinates} respectively. The orthonormal basis vectors for the parametrization are 
\(\That = \ds \boldsymbol{r}(s)\),
\(\Nhat = \ds \That/\kappa(s)\),
and 
\(\Bhat = \That \times \Nhat\),
which correspond to the \emph{tangential}, \emph{normal} and \emph{binormal} direction as illustrated in \cref{fig:JSC_example}. We refer to \(\kappa(s)\) as the \emph{curvature function}.

\begin{figure}
   \centering\includegraphics[width = 0.925\linewidth]{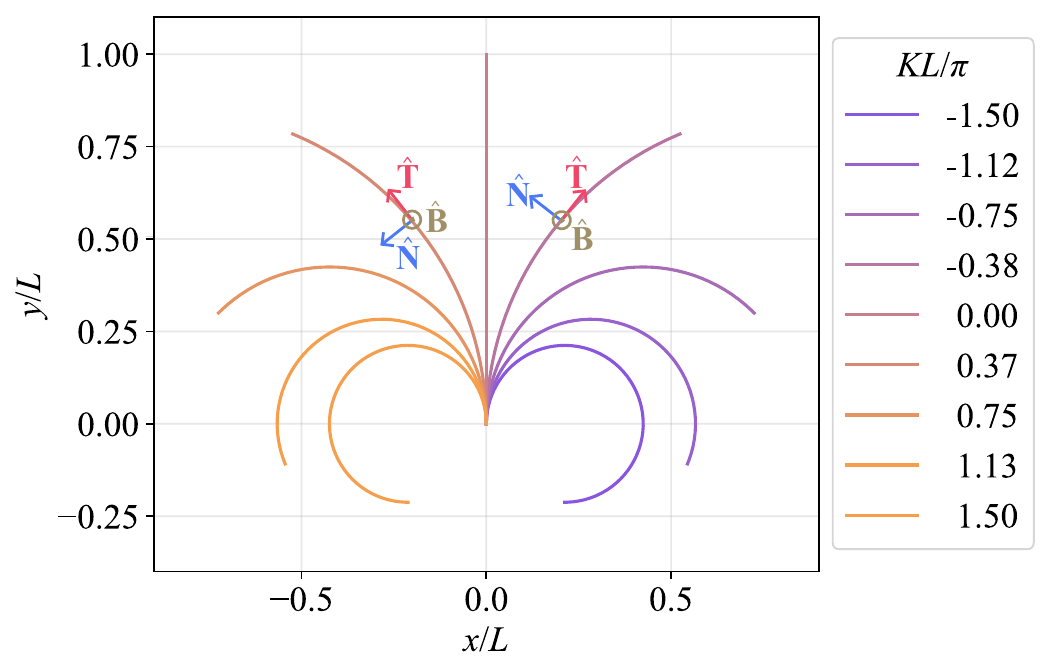}
   \caption{Lines of constant curvature as a function of the curvature amplitude \(K\). Changing the sign of \(K\) from positive to negative reverses the direction of the curvature from anti-clockwise to clockwise.
   The basis vectors, as given in Eqn.~\ref{eq:unit_vectors}, are drawn for two of the curves.}
   \label{fig:curve_sketches}
\end{figure}

The basis vectors obey a Frenet-Serret like equation 
\begin{equation}
    \begin{pmatrix}
        \ds \That \\
        \ds \Nhat \\
        \ds \Bhat
    \end{pmatrix}
    =
    \begin{pmatrix}
        0 & \kappa(s) & 0 \\
        -\kappa(s) & 0 & 0 \\
        0 & 0 & 0
    \end{pmatrix}
    \begin{pmatrix}
        \That \\
        \Nhat \\
        \Bhat
    \end{pmatrix},
\end{equation}
and are accompanied by a non-trivial metric tensor \cite{salamone2022curvature, Ortix2015} 
\(G_{\mu\nu} = \text{diag}(\eta^{2}, 1, 1)\),
with
\(\eta(s, n) = 1 - \kappa(s) n\). When the system is in equilibrium, it is sufficient to consider only the retarded propagator, \(\gR\). Moreover, in the 1D transport limit, where \(n, b \rightarrow 0\) with \(\ds \kappa(s)\) finite, the non-zero Christoffel symbols do not contribute to the Usadel equation, and the basis vectors remain orthonormal, such that the Usadel equation for the retarded propagator takes the form \cite{salamone2022curvature}
\begin{equation}\label{eq:usadel_gr}
    D_{F}\dst ( \gR \dst \gR) + i \comm{\epsilon \that_3 + \hat{\Sigma}}{\gR} = 0.
\end{equation}
In this form, the effect of the curvature is encoded in the rotation of the basis vectors (and Pauli matrices) along the arclength. 

Considering a superconductor/ferromagnet heterostructure, we take the self-energy
\(\hat{\Sigma} = \hat{\Delta}\) on the superconducting side, and \(\hat{\Sigma} = h^{\mu}(s)\text{diag}\left[\sigma_{\mu}(s), \sigma_{\mu}^{*}(s)\right]\) on the ferromagnetic side. 
\(h^{\mu}(s)\) is the exchange field and  \(\sigma_{\mu}(s)\) the Pauli matrices along the \(\mu\)-direction,
and 
\(\hat{\Delta} = \text{antidiag}(\Delta, -\Delta, \Delta^{*}, -\Delta^{*})\), where \(\Delta\)
is the superconducting order parameter. In the following, we keep $\Delta$ fixed and equal to its value in an infinite superconductor, thereby ignoring the inverse proximity effect caused by the interface. All lengths are scaled relative to the diffusive superconducting coherence length, denoted by $\xi$, which is typically of the order of a few tens of nanometers.

For the boundary conditions we will employ the Kupryanov-Lukichev boundary conditions \cite{kuprianov1988influence}, which take the form
\begin{equation}\label{eq:KL_unparam}
    \gc_{j} \dtilde_{I} \gc_{j} = \frac{1}{2 L_{j} \zeta_{j}} [\gc_{L}, \gc_{R}],
\end{equation}
where the index \(j = \{L, R\}\) refers to the left and right side of the interface, $L_{j}$ is the length of material, $\dtilde_{I}$ is the gauge covariant derivative at the interface, and $\zeta_{j}=R_B/R_j$ quantifies the interfacial resistance via the ratio of barrier resistance $R_B$ to bulk resistance $R_j$. 

\subsection{Parametrization of non-constant, step-like curvature functions}\label{Sec:Non-const}
We consider junctions consisting of straight and curved parts in series. Mathematically, this is done by considering curvature functions, \(\kappa(s)\), consisting of a series of step-functions, as opposed to constants for constant curvature \cite{Salamone2021,salamone2022curvature}. We will demonstrate that approximate step-functions can be rigorously included in the formalism and use them to define the J-, C- and S-classes of curves.

We will consider planar curves that can be parameterized using curvature functions of the form 
\begin{equation}
    \kappa(s) = K \sum_{j} \alpha_{j} H(s - q_{j}),
\end{equation}
where we refer to \(K\) as the \emph{curvature amplitude}, and \(H\) is the Heaviside step function.
Particular values from the sets \(\{q_{j}\}\) and \(\alpha_{j} = \pm 1\) determine where the curvature is present and/or reversed. The orientation of the unit vectors relative to their initial configuration, \(\theta(s)\), can then be determined from the integral of the curvature function \(\kappa(s)\):
\begin{align}
        \theta(s) &= \int_{\lambda = 0 }^{\lambda = s} \kappa(\lambda; K, q) d\lambda \nonumber \\
    &= K \sum_{j} \alpha_{j} H(s - q_{j}) \cdot (s - q_{j}),
    \label{eq:theta_s}
\end{align}
such that we may parameterize the basis vectors as 
\begin{subequations}
\label{eq:unit_vectors}
\begin{align}
    \That &=
    - \sin \theta (s) \hat{x} + \cos \theta (s) \hat{y}, \\
    \Nhat &= - \cos \theta (s) \hat{x} - \sin \theta (s) \hat{y} \\
    \Bhat &= \hat{z}.
\end{align}
\end{subequations}
The conceptually simple case of constant curvature, with \(\kappa(s) = K\), is shown in \cref{fig:curve_sketches}. 
When we have fixed the binormal basis vector, \(\Bhat\), positive (negative) curvature functions \(\kappa(s)\) correspond to counter-clockwise (clockwise) rotation.

For this demonstration of non-constant curvature, we consider three conceptually simple cases: wires curved at one end (J-type), curved in the same direction at each end (C-type), or curved in opposite directions on each end (S-type), as shown in \cref{fig:JSC_example}. The associated curvature functions are 
\begin{subequations}
\begin{eqnarray}
    \kappa^{J}(s; K, q) &=& K H(s - q), \label{Eq:kappaj}\\
    \kappa^{C}(s; K, q) &=& 
    \begin{cases}
        K \quad \hspace{2mm}\text{if } s < q \\
        0 \quad \hspace{3.5mm}\text{if } q < s < L - q \\
        K \quad \hspace{2mm}\text{otherwise,}\\
    \end{cases}\\
    \kappa^{S}(s; K, q) &=& 
    \begin{cases}
        K \quad \hspace{2.7mm}\text{if } s < q \\
        0 \quad \hspace{4mm}\text{if } q < s < L - q \\
        -K \quad \text{otherwise.}\\
    \end{cases}
\end{eqnarray}
\end{subequations}
Using \cref{eq:theta_s}, this gives
\begin{subequations}
\begin{align}
    \theta^{J} (s; K, q) &= 
    \begin{cases}
        Ks \quad \hspace{15mm}\text{if } s < q \\
        K q \quad \hspace{15mm}\text{otherwise,}
    \end{cases}\\
    \theta^{C} (s; K, q) &= 
    \begin{cases}
        Ks \quad \hspace{15mm}\text{if } s < q \\
        K q \quad \hspace{15mm}\text{if } q < s < L - q\\
        K (s \!+\! 2q \!-\! L) \quad \text{otherwise,}
    \end{cases} \hspace{-2mm}\label{eq:theta_c}\\
    \theta^{S} (s; K, q) &= 
    \begin{cases}
        Ks \quad \hspace{15mm}\text{if } s < q\\
        K q \quad \hspace{15mm}\text{if } q < s < L - q\\
        K (L - s) \quad \hspace{5.5mm}\text{otherwise.}
    \end{cases}
\end{align}
\end{subequations}
Here \(L\) is the length of the parameterized section, and we refer to \(q\)
as the \emph{shape parameter}. 

To ensure continuously varying observables, and to rigorously derive the Usadel equation in 1D, we require a continuous \(\kappa(s)\) and finite \(\ds \kappa(s)\). To achieve this we may approximate the Heaviside function by the hyperbolic tangent, and thus the curvature functions may, as an example, take the form
\begin{equation}
    \label{eq:tanh_kappa}
    \kappa^{J}(s; K, a, q) = K\left(1 - \frac{e^{a(s-q)}}{1 + e^{a(s-q)}}\right).
\end{equation}
Here we have defined a \emph{sharpness parameter} \(a\), indicating the abruptness of the change in curvature. In the limit \(a \rightarrow \infty\) we retrieve the step-function, giving \cref{Eq:kappaj}. 
We may further insert \cref{eq:tanh_kappa} into \cref{eq:theta_s}, to get 
\begin{equation}
    \label{eq:tanh_tetha}
    \theta^{J}(s; K, a, q) = K \left[ s + \frac{1}{a} \ln \left(\frac{1 + e^{-aq}}{1 + e^{a(s-q)}}\right) \right].
\end{equation}

There are closed form expressions for $\theta^{S,C}$, 
as well as for more generalized cases. There is, however, in general no closed form expression for the parameterized curve itself.

\subsection{Numerical approach: Riccati parametrization}\label{Sec:Riccati}
To numerically solve \cref{eq:usadel_gr}, it is helpful to employ a more tractable parameterization of the quasiclassical Green's function. We use the Riccati parametrization \cite{Jacobsen2015b,SchopohlMaki1995}:
\begin{equation}
    \label{eq:Riccati_GR}
    \gR = 
    \begin{pmatrix}
        N(1 + \gamma \gt) & 2N\gamma \\
        -2\Nt \gt & -\Nt(1 + \gt \gamma)
    \end{pmatrix},
\end{equation}
with \(N = (1 - \gamma \gt)^{-1}\) and \(\gt(s, \varepsilon) = \gamma^{*}(s, -\varepsilon)\), i.e. tilde-conjugation implies inversion of the quasiparticle energy and element wise complex conjugation. The parametrization reduces the dimensionality of the problem from an equation of \((4\times4)\)-matrices to coupled equations for \((2\times2)\)-matrices, and bounds the norm of the matrix elements $\in[0,1]$. 

In the ferromagnetic region, the Usadel equation is then parametrized as 
\begin{equation}
    \label{eq:Riccati_Usadel}
    D_{F} \!\left\{\!\ds \gamma \!+\! 2(\ds \gamma) \Nt \gt (\ds \gamma)\!\right\} \!=\! -2i\varepsilon\gamma - i h^{\mu} [\sigma_{\mu}(s) \gamma - \gamma \sigma_{\mu}^{*}(s)],
\end{equation}
with the index $\mu=\{\hat{T},\hat{N},\hat{B}\}$, and \(\sigma_{\mu}(s)\) are the corresponding Pauli matrices.
The unit vectors \(\That, \Nhat, \Bhat\) are determined from \cref{eq:unit_vectors} with \(\theta(s)\) being, e.g., \cref{eq:tanh_tetha} for a J-type curve or \cref{eq:theta_c} for a C-type curve. 
Inserting the Riccati parametrization \cref{eq:Riccati_GR} into the boundary conditions \cref{eq:KL_unparam} at the superconductor-ferromagnet interface, the boundary conditions for the matrix $\gamma$ become \cite{Jacobsen2015b,salamone2022curvature}:
\begin{subequations}
\label{eq:KLbc}
\begin{eqnarray}
    \dtilde_{I}\gamma_S &=& \frac{1}{L_S\zeta_S}(1\!-\!\gamma_S\gt_F)N_F(\gamma_F\!-\!\gamma_S), \label{eq:KLbc1} \\
    \dtilde_{I}\gamma_F &=& \frac{1}{L_F\zeta_F}(1\!-\!\gamma_F\gt_S)N_S(\gamma_F\!-\!\gamma_S). \label{eq:KLbc2}
\end{eqnarray}
\end{subequations}
The corresponding equations for $\gt$ are obtained by tilde conjugation of \cref{eq:Riccati_Usadel,eq:KLbc1,eq:KLbc2}.

We can then solve the complete, Riccati parameterized Usadel equation \eqref{eq:Riccati_Usadel} in a 1D ferromagnetic region with either constant or non-constant curvature functions, coupled to bulk \(s\)-wave superconductors, employing the Kupriyanov-Lukichev boundary conditions \eqref{eq:KLbc} at the interfaces.
The specific implementation is based on \cite{GENEUS2018}, modified to accommodate the non-constant curvature as specified here, and run using the \texttt{bvp6c} package for \textsc{matlab}.

\section{Results}\label{sec:Results}
We begin by examining the Usadel equations in the limit of weak proximity coupling in Sec.~\ref{Sec:WPE}, where we can get analytic insight into the dominant mechanisms influencing the spin transport. We discuss this analytic insight in the context of the simpler J-type curve in Sec.~\ref{Sec:J}. We then present numerical results for the full proximity effect in C- and S-type junctions. In particular, we show a chirality-dependent switch in the magnetization and spin current density of C-type curves in Sec.~\ref{Sec:Magnetization}. We compare the equilibrium current of C- and S-type curves in Sec.~\ref{Sec:EquiCurrent}, and show that only C-type undergo a $0-\pi$ transition, since the combination of \textit{equal} chiralities in an S-type junction interferes destructively.

\subsection{Weak proximity equations}\label{Sec:WPE}
In the weak proximity limit, we assume the anomalous Green's function to be small, such that 
\begin{equation}\label{Eqn:gR}
    \gR = \begin{pmatrix}
        1 & f \\
        -\Tilde{f} & 1
    \end{pmatrix},
\end{equation}
and parametrize the anomalous Green's function using the d-vector formalism:
\begin{equation}
    \label{eq:d-vec}
    f = (f_{0} + d_{\mu}\sigma_{\mu}) i \sigma_{y}.
\end{equation}
Here \(f_{0}\) and \(d_{\mu}\) are the singlet and triplet amplitudes respectively. In this limit, we need only to consider the terms of the Riccati parametrized equation that are linear in \(\gamma\), and the equation takes the following form in the ferromagnetic region:
\begin{subequations}
\begin{eqnarray}
    \frac{iD_{F}}{2}\ds^{2}f_{0} \!&=&\! \varepsilon f_{0} + h_{\mu}d_{\mu}, \label{eq:WPE_f0} \\
    \frac{iD_{F}}{2}\ds^{2}d_{T} \!&-&\! iD_{F}(\ds\kappa \!+\! \kappa \ds)d_{N} \!=\!
    \left[ \varepsilon \!+\! \frac{iD_{F}\kappa^{2}}{2} \right]\! d_{T} \!+\! h_{T}f_{0}, \nonumber\\ \label{eq:WPE_dT} \\
    \frac{iD_{F}}{2}\ds^{2}d_{N} \!&+&\! iD_{F}(\ds\kappa \!+\! \kappa \ds)d_{T} \!=\!
    \left[ \varepsilon \!+\! \frac{iD_{F}\kappa^{2}}{2} \right]\! d_{N} \!+\! h_{N}f_{0}, \nonumber\\ \label{eq:WPE_dN}\\
    \frac{iD_{F}}{2}\ds^{2}d_{B} \!&=&\! \varepsilon h_{B} + h_{B}f_{0}.\label{eq:WPE_dB}
\end{eqnarray}
\end{subequations}
With a non-zero \(\kappa\), the triplet components undergo spin precession and spin relaxation. We identify the precession with the terms having a first order derivative, while the imaginary contributions to the energy represent the spin relaxation and loss of spin information from impurity scattering. While the curvature $\kappa(s)$ provides a mechanism for rotating between different triplet components, generating the spin-polarized triplets that are robust in magnetic fields (the so-called long-ranged component), it also induces spin relaxation. Having a non-constant curvature function, \(\kappa(s)\), enables optimization of the long-range triplet generation and retention. 

The triplet component with zero spin projection is short-ranged in a magnet, with d-vector parallel to the exchange field; the triplet component with spin polarization along the exchange field is long-ranged, with a d-vector perpendicular to the exchange field. For example, if the exchange field is directed along $\hat{T}$, the short-ranged triplets can be identified with $d_T$, while $d_N$ and $d_B$ are long ranged.

In the weak proximity equations, we can see a curvature dependent mixing between \(d_{N}\) and \(d_{T}\). The mixing depends on \(\ds \kappa\), which couples to the amplitudes themselves, \(d_{\mu}\), rather than their derivatives. That is, the mixing is enhanced in regions where the curvature changes abruptly. Moreover, it is possible to change the sign of \(\ds \kappa\) without changing that of \(\kappa\).

\subsection{J-type ferromagnets and chirality-dependent spin polarization}\label{Sec:J}
\begin{figure}
    \centering
    \includegraphics[width=0.925\linewidth]{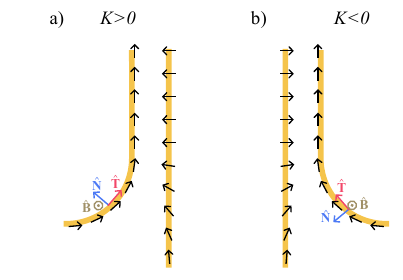}
    \caption{Illustration of the exchange field, i.e. the tangent vectors, for a J-like curve at equidistant points along the curve for (a) \(K > 0\) and (b) \(K < 0\). An equivalent system with a straight wire and a rotating exchange field is also shown for both cases. The arrows indicate the direction of the exchange field, and are placed with equidistant (center-to-center) spacing for both cases. For a curvature amplitude of $KL=\pi/2$, the exchange field in the corresponding straight system is oppositely aligned at the end of the wire.}.
    \label{fig:unravel}
\end{figure}

The simplest class of wires with non-constant curvature is the J-type ferromagnet (see \cref{fig:JSC_example}(a) and \cref{fig:unravel}). For any superconductor-ferromagnet (SF) bilayer, the ferromagnet's exchange field will convert a proportion of the superconducting singlet correlations into triplets with non-zero spin projection (short-ranged triplets), which is clear from \cref{eq:WPE_f0}. The curvature is then responsible for the rotation of the triplet vector, converting between tripets with and without spin projection. Since the short ranged triplet correlations decay rapidly in a ferromagnet, the best way of preserving long-ranged, spin-polarized triplets in the system is to have a region of sharp curvature near the superconducting interface for rapid conversion from short-ranged to long-ranged, and then a region of zero curvature, to avoid rotating the long-ranged components back into short-ranged triplets \cite{Salamone2021}. Although the conversion mechanism is in this case provided by the curvature, the process governing interconversion between components is the same as for straight multilayers with magnetic misalignment \cite{HouzetBuzdin2007}. However, real-space curvature provides new mechanisms for design of variable exchange-field misalignment, and dynamic control of their relationship. 

The utility of curvature for controlling the generation of spin-polarized superconducting correlations becomes particularly clear when considering the equivalent field of two J-curves with opposite chirality, as in \cref{fig:unravel}.
We see that the sign of the curvature at the superconducting interface governs the direction of the exchange field at the edge of the corresponding straight wire, which will be opposite for a curvature amplitude of $KL = \pi/2$. Transport through such wires will therefore experience opposite chiralities, and we can detect signatures of this chirality in the observables for superconducting correlations.

When we have a superconductor-ferromagnet-superconductor (SFS) junction instead of a bilayer, we must take into account the triplet conversion from both superconductors. Since the proximity effect at each interface will experience a chiral-dependent triplet conversion, combinations of different chiralities can therefore be expected to interfere constructively or destructively along the wire, which we explore below.

\subsection{C- and S-type junctions}\label{Sec:CS}
With our knowledge of the chirality-dependent polarization of a J-type bilayer, we can now examine and contrast the effects of combining different chiralities in SFS Josephson junctions, as illustrated by the C- and S-type wires in \cref{fig:JSC_example}. In terms of the proximity effect, a C-type junction combines \textit{opposite} chiralities with respect to the interface of each superconductor, and the S-type combines \textit{equal} chiralities. That is, curving the regions close to the interface rotates the exchange field to give a component perpendicular to the direction of transport in the equivalent straight case (see \cref{fig:unravel}). For the C-type junction, this component points in the opposite direction near each interface, and points in the same direction for S-type. Limiting cases now give the corresponding exchange field profiles of a junction with three misaligned ferromagnets \cite{HouzetBuzdin2007,Volkov2003,Volkov2010,Trifunovic2010}, or a ferromagnet with domain walls at the interfaces \cite{Bergeret2001}. As before, sharp curvatures at a superconducting interface promotes long-ranged triplet generation, and any further rotation along the wire will induce spin relaxation. Parametrization with the shape parameter, \(q\), lets us vary the relative length of the straight and curved  segments, and allows for separation of curvature-induced and length-induced effects.

We adopt the convention that with \(K>0\), we have \(\kappa(s=0) > 0\), and the wire curves counter-clockwise at the origin (as in \cref{fig:curve_sketches}). How the proximity-induced triplets will combine in a junction will therefore depend on the radius of curvature, and any change in sign of $K$ along the arclength.
Below, we will show that the full Usadel equation predicts a chirality-dependent magnetization and spin current density in C-type junctions with constant curvature in Sec.~\ref{Sec:Magnetization}, and go on to show the effect of constructive and destructive chirality combinations in the critical current of C- and S-type junctions in Sec.~\ref{Sec:EquiCurrent}.

\subsubsection{Chiral signatures:\\ Magnetization and spin current density}\label{Sec:Magnetization}
Once we have determined the Green's function, we may determine the equilibrium spin accumulation, or proximity-induced magnetization, \(M_{\mu}\) in the wire in the \(\mu\) direction from \cite{Champel2005effect}
\begin{equation}
    M_{\mu} = M_{0} \int_{-\infty}^{\infty} d\varepsilon
    \Tr \left\{\text{diag}(\sigma_\mu, \sigma^{*}_\mu) \gK \right\},
\end{equation}
where we use the ansatz 
\(\gK = (\gR - \gA) \tanh{\frac{\beta\varepsilon}{2}}\) in equilibrium. The coefficient \(M_{0} = g \mu_{B}N_{0}\Delta / 16\), with the Landé \(g\)-factor \(g\approx2\) for electrons, \(\mu_{B}\) is the Bohr magneton, $\beta=\frac{1}{k_{B}T}$, and \(N_{0}\) is the normal-state density of states at the Fermi-level. 
\begin{figure*}
    \centering
    \includegraphics[width = 0.925\linewidth]{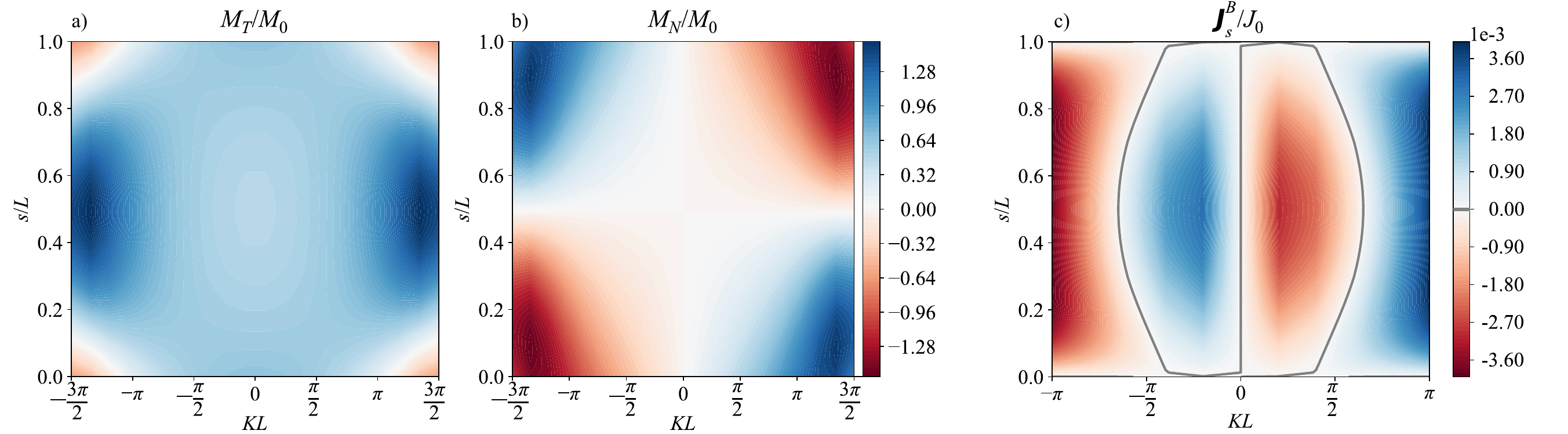}
    \caption{Chirality-dependent signatures. Contour plots of the magnetization in the (a) tangential, \(M_{T}\) and (b) normal, \(M_{N}\) direction for an SFS junction of length \(L = 0.8\xi\), with an exchange field of \(\boldsymbol{h} = \Delta \hat{T}\) as a function of arclength and constant curvature with amplitude \(KL\). (c) Spin current density \(\boldsymbol{J_{s}^{B}}\) as a function of normalized position \(s/L\) for different constant curvature amplitudes \(KL\) for SFS junctions with \(L = 2\xi, \phi = \pi/2, T = 0.005T_{}c\)  and \(\boldsymbol{h} = \Delta \hat{T}\). The contour with \(\boldsymbol{J_{s}^{B}} = 0\) is traced with a thick gray line.}
    \label{fig:L2Magnetization_contour}
\end{figure*}

To demonstrate the chiral signatures, we begin by considering the case of constant curvature in C-type junctions, with different chiralities/sign of $K$, as depicted in \cref{fig:curve_sketches}. We show the induced magnetization along a wire for $K \in (-3\pi/2, 3\pi/2)$ in \cref{fig:L2Magnetization_contour}, and see that reversing the direction of the curvature also reverses the sign of the magnetization in the normal direction in \cref{fig:L2Magnetization_contour}(b), while the tangential magnetization [\cref{fig:L2Magnetization_contour}(a)] is unchanged. This leaves a chirality-dependent, observable signature of the spin polarization.
We may understand this in terms of a straight SFS-junction with a rotating exchange field, which is equivalent to a curved wire with a tangential exchange field. The curved wire and its equivalent straight exchange field is illustrated in \cref{fig:unravel}. In the straight case, reversing the chirality of the exchange field rotation would reverse the spin accumulation orthogonal to the transport direction. Note that the parameter determining the chirality, \(K\), can vary continuously, and that the response, \(M_{T,N}(s)\) varies continuously as a function of \(K\), such that this still holds in the limit \(K \rightarrow 0\). 

The effect of the exchange field chirality also gives a signature in the spin current density \(\boldsymbol{J}_{s}^{\mu}\), which is given in the \(\mu\)-direction by the Keldysh component 
\begin{equation}
    \label{eq:spin_current_density}
    \boldsymbol{J}_{s}^{\mu} = 
    \frac{\hbar N_{0} D}{4} \int_{-\infty}^{\infty} d\varepsilon\Tr\left\{
    \text{diag}(\sigma_\mu, \sigma^{*}_\mu) \that_{3} \left(\gc\ds\gc\right)^{K}
    \right\}.
\end{equation}
In equilibrium, we can see the contributions of the triplet components in weak proximity in the binormal component:
\begin{eqnarray}
    \boldsymbol{J}_{s}^{B} &\!=\! 
    &-\frac{\hbar N_{0} D}{4} \!\!\int_{-\infty}^{\infty}\!\!\!
    d\varepsilon \text{Im}
    \left[
    d_{T} \ds \tilde{d}_{N} \!-\! \tilde{d}_{N} \ds {d}_{T} \!+\! \text{t.c.}\!
    \right]\!\tanh{\frac{\beta\varepsilon}{2}}\nonumber\\
    &&-\kappa(s)\frac{\hbar N_{0} D}{2} \!\!\int_{-\infty}^{\infty}\!\!\!
    d\varepsilon \text{Im}
    \left[
    d_{T}\tilde{d}_{T} \!+\! {d}_{N}\tilde{d}_{N}\!
    \right]\!\tanh{\frac{\beta\varepsilon}{2}},\label{Eqn:JBwpe}
\end{eqnarray}
where t.c. indicates the tilde conjugate of the preceding terms. The normal and tangential contributions to the spin current density are zero, since the terms couple to $d_B$ and the derivative of $\sigma_B$, which give zero contribution. 

By inspecting the weak proximity expression \eqref{Eqn:JBwpe}, we can infer that the spin current density should also change sign under reversal of K: the second term is directly odd in $\kappa$, and we saw from the magnetization that reversing K reverses $d_N$ and $\tilde{d}_N$, which will reverse the sign of the first term. This is shown in \cref{fig:L2Magnetization_contour}(c), for a phase difference of $\phi=\pi/2$, where the triplet contribution is maximal. We also see a second sign change for higher curvatures, here around $KL/\pi\approx 0.6$, due to a \(0-\pi\) transition \cite{Salamone2021}. 

\subsubsection{Mixed chiralities in the critical charge current and ground state transitions}\label{Sec:EquiCurrent}
We can consider the combination of mixed chiralities with respect to the local superconducting interface in SFS junctions by comparing the behaviour of C- and S-type junctions with varying degrees of curvature, and will find that the triplet response to the changing spin quantization axis will interfere either constructively or destructively. 

\begin{figure}
    \centering
    \includegraphics[width = 0.925\linewidth]{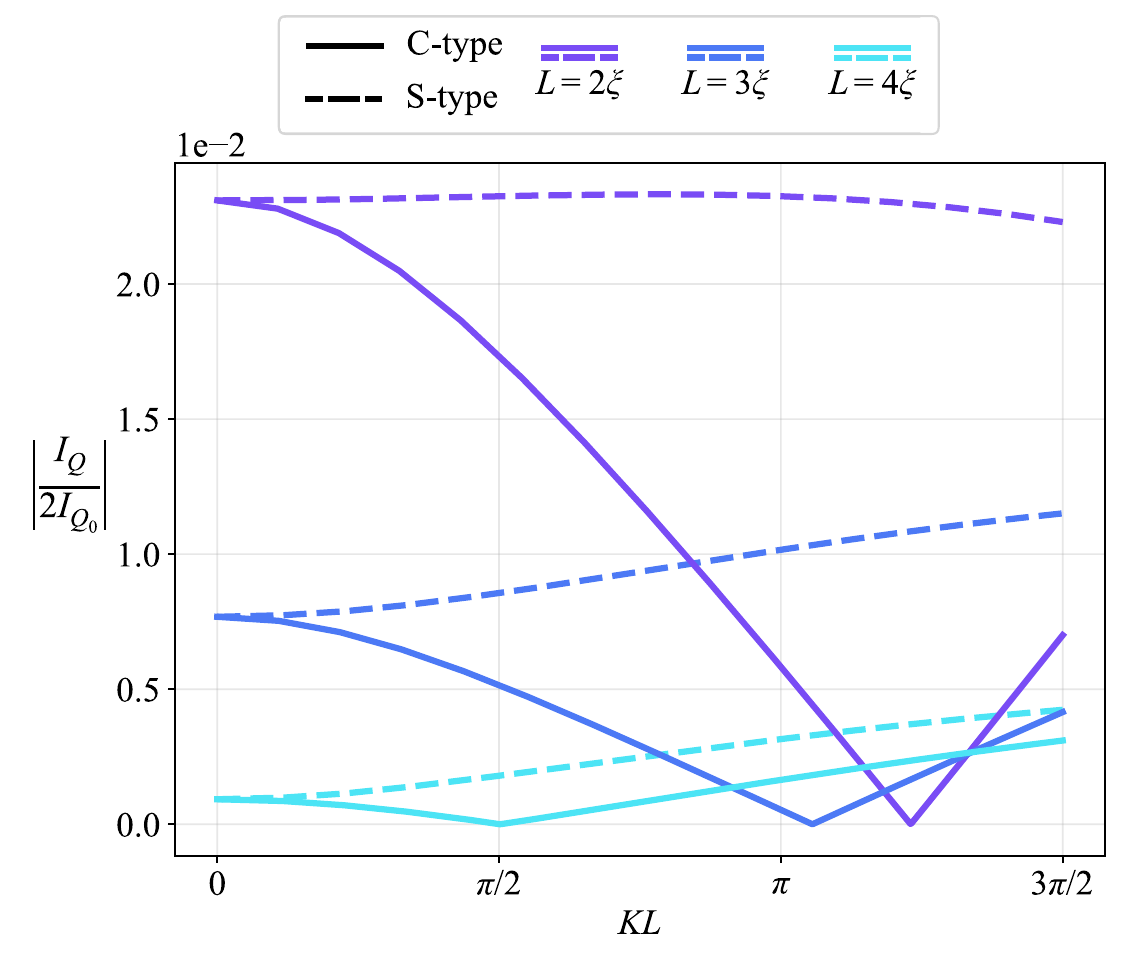}
    \caption{Magnitude of the critical current as a function of the curvature amplitude \(K\) for different lengths \(L\) of C-type (solid line) and S-type (dashed line) with \(q/L = 0.3\) junctions. \(T = 0.005 T_{c}\), \(\vec{h} = \Delta \hat{T}\), and \(\zeta = 3\). The C-type junctions display a curvature induced \(0-\pi\) transition, while the S-type has none.} 
    \label{fig:current}
\end{figure}

The curvature-induced \(0-\pi\) transition for in-plane SFS junctions with constant curvature is due to the singlet-triplet conversion~\cite{Salamone2021}. Here, we will compare the behaviour of junctions with mixed chiralities at the interfaces. The charge current in an SFS junction is given by the Keldysh component 
\begin{equation}
    \frac{I_{Q}}{I_{Q_{0}}} = \int_{-\infty}^{\infty}
    d\varepsilon \Tr \left( \that_{3}\gc \ds \gc\right)^{K},
\end{equation}
which, in equilibrium, simplifies to
\begin{equation}
    \frac{I_{Q}}{I_{Q_{0}}} \!=\! \int_{-\infty}^{\infty}\!\!\!
    d\varepsilon \Tr\left\{
    \that_{3}\left( \gR \ds \gR \!-\! \gA \ds \gA \right)
    \right\}\tanh{\frac{\beta\varepsilon}{2}}.
\end{equation}
Here \(I_{Q_{0}} = N_{0}eD_{F}A\Delta_{0}/4L\), where $e$ is the electron charge, \(A\) the cross sectional area of the wire, and \(\Delta_{0}\) is the gap of the two identical bulk-like superconductors. 
The magnitude of the critical current is given in \cref{fig:current} as a function of the curvature amplitude \(K\), for different ferromagnet lengths \(L\) and shapes (C, S). It is clear that here the C-type curves have a curvature-induced \(0-\pi\) transition, whereas the S-type curves only allow for a modulation of the current.

In the case of a straight ferromagnet, the \(0-\pi\) transition is governed by the length of the ferromagnet, due to the modulation in the acquired phase difference between the correlated spins \cite{Kontos2002,Buzdin2005}. This length-based phase acquisition is modified by the phases acquired due to curvature-induced rotation, but the length will clearly still be a factor in controlling the ground state. For example, for the parameters chosen in \cref{fig:current}, longer junctions (e.g. $L=6\xi$) would already be in the $\pi$ ground state for $K=0$, with initial reversal of the sign of the critical current. In that case, the $S$-type would display a $\pi-0$ transition, whereas the $C$-type would remain in the $\pi$-state.
We can compare the singlet and triplet contributions to the current by decomposing the anomalous Green's function using \cref{eq:d-vec} and writing \(I_{Q}/I_{Q_0} = I_{0} + I_{\mu} + I_{\kappa}\), with
\begin{subequations}
\begin{eqnarray}
    \frac{I_{0}}{I_{Q_{0}}} &=& -8\!\! \int_{0}^{\infty}\!\!\! d\varepsilon \text{Re}\left\{
    \Tilde{f}_{0}\ds f_{0} \!-\! f_{0}\ds \Tilde{f}_{0}
    \right\}\!\tanh{\!\frac{\beta\varepsilon}{2}},\\
        \frac{I_{\mu}}{I_{Q_{0}}} &=& +8\!\! \int_{0}^{\infty}\!\!\! d\varepsilon \text{Re}\left\{
    \Tilde{d}_{\mu}\ds d_{\mu} \!-\! d_{\mu}\ds \Tilde{d}_{\mu}
    \right\}\!\tanh{\!\frac{\beta\varepsilon}{2}},\\
        \frac{I_{\kappa}}{I_{Q_{0}}} &=& 16 \kappa(s) \!\!\int_{0}^{\infty}\!\!\!d\varepsilon \text{Re}\left\{
    \Tilde{d}_{N}d_{T} \!-\! \Tilde{d_{T}}{d}_{N}
    \right\}\!\tanh{\!\frac{\beta\varepsilon}{2}}.
\end{eqnarray}
\end{subequations}
Here $I_0$ represents the singlet contribution, $I_\mu$ is the triplet contribution in the $\mu$ direction, and $I_\kappa$ is the inverse Edelstein contribution, present when the d-vector rotates \cite{Amundsen2017}. That is, there will be no inverse Edelstein contribution in a straight segment of the wire here.

\begin{figure}
    \centering
    \includegraphics[width = 0.925\linewidth]{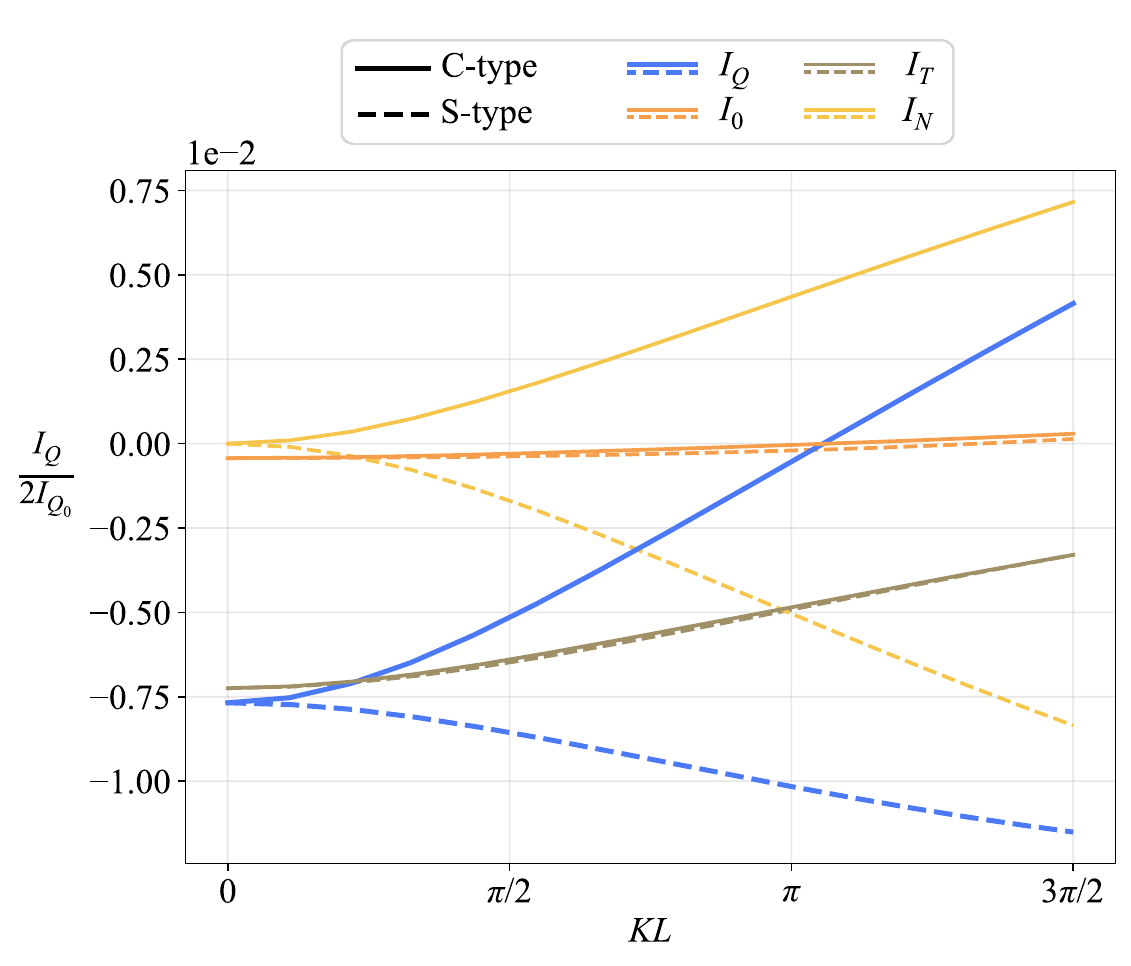}
    \caption{The critical charge current for a C- and S-type junction, as a function of the normalized curvature amplitude \(KL\), at \(s = L/2\) and with \(q/L = 0.3\). The current is separated into the total, \(I_{Q}\), singlet \(I_{0}\), and short and long range triplet contributions, \(I_{T}\) and \(I_{N}\) respectively. Here \(L = 3 \xi, T = 0.005T_{c}\), \(\boldsymbol{h} = \Delta \hat{T}, \zeta = 3\) and \(\phi = \pi/2\). The C-type junction undergoes a \(0-\pi\)-transition at \(KL \sim \pi\).}
    \label{fig:current_components}
\end{figure}
As discussed above, the curvature induces generation of the long range triplet correlations by rotating the $d$-vector from the short ranged correlations, giving the long-range current component \(I_{N}\). The first thing to notice about \cref{fig:current_components} is that $I_N$ is almost antisymmetric with respect to the geometry being either of the S or the C type. We can understand this by employing the construction discussed in \cref{fig:unravel}. By deforming the curved wire to a straight equivalent, while keeping the exchange field orientation fixed, one finds that for C-type geometries, the effective orthogonal exchange field (i.e. the component along the normal direction evaluated in the middle of the junction, $\mathbf{h}\cdot \hat{N}(s=L/2)$) points in opposite directions at the ends, whereas for S-type geometries, it points in the same direction. In the former case, the $d_N$ triplets generated at the ends, are expected to have a phase difference of $\pi$, which comes in addition to the Josephson effect, due to the sign change of $h_N$. This produces a current contribution $I_N$ which flows in the opposite direction to the latter case, where there is no such sign-change. 
As we increase the curvature amplitude \(K\), the C-type geometry therefore features a \(0-\pi\) transition in the charge current $I_Q$, which is shown in Fig. \ref{fig:current_components}. 

\section{Triplet-SQUID design and discussion}\label{sec:Discussion}
A triplet-SQUID design has recently been proposed for use in a superconducting memory \cite{Glick2018}. It uses the conventional singlet-triplet conversion mechanism of misaligned magnetic multilayers. The two SQUID weak links are comprised of magnetic trilayers, and the magnetization of the central layer of one link can be switched at lower field than the other. They use this to demonstrate that the ground state of a trilayer Josephson junction can switch between $0$ and $\pi$ depending on the relative magnetization of the layers \cite{HouzetBuzdin2007,Volkov2003,Volkov2010,Trifunovic2010}.  

In our case, both C- and S-type curves give triplets in the middle of the ferromagnetic weak link, but only C-type gives a $0-\pi$ transition. This means we can achieve the same triplet-SQUID behaviour by including a ferromagnetic weak link of different curvature class in each arm of the SQUID. That is, we can use a single ferromagnetic layer for each weak link, instead of magnetic multilayer structures that are difficult to control and model accurately. Instead of needing a different sample for each new sample thickness, this design can also be parallelized by varying the planar length in parallel on the same sample. In addition, it will now be possible to switch the ground state of the junctions in-situ by employing strain, for example with electrical control via a piezoelectric substrate

In this work, we have focused on identifying signatures of different chiralities of curvilinear magnets at the interface with conventional superconductors. We chose an arbitrary shape parameter $q$ for these demonstrations, but it would be useful to optimize the shape for maximal critical current of a junction of fixed arclength, for example.

More generally, we have shown that geometric curvature in a magnet is a versatile tool for designing bespoke effective spin-orbit coupling profiles that can vary throughout the structure. We have shown how this can be harnessed in superconducting proximity heterostructures, to control the spin polarization of superconducting correlations, giving chiral signatures in the magnetization and spin current density. When heterostructures contain multiple superconducting interfaces, the chirality at each interface should be considered in combination with the chiralities of the others. By combining elements of different chirality, as demonstrated by comparing C- and S-type junctions, we will have different mechanisms to control the ground state of the junction. We can implement this directly for improved functionality in devices, and we show how a combination of C- and S-type junctions can reproduce magnetic multi-layer triplet-SQUID behaviour.

\begin{acknowledgments}
The computations have been preformed on the SAGA supercomputer provided by UNINETT Sigma2 ---
the National Infrastructure for High
Performance Computing and Data Storage in Norway. We acknowledge funding via the “Outstanding Academic Fellows”
programme at NTNU, the "Sustainable Development Initiative" at UiO, the Research Council of Norway Grant
No. 302315, as well as through its Centres of Excellence
funding scheme, Project No. 262633, “Center for Quantum
Spintronics”. MA acknowledges support from the Swedish Research Council (Grant No. VR 2019-04735 of Vladimir Juri\v{c}i\'c). Nordita is supported in part by NordForsk.
\end{acknowledgments}

\bibliography{main}

\end{document}